\documentclass[conference,a4paper]{APSIPA2025}
\usepackage{amsmath,amssymb,amsfonts}
\usepackage{graphicx}
\usepackage{multirow}
\usepackage{threeparttable}

\usepackage{xcolor}
\usepackage{balance}
\usepackage{multirow, bm}
\usepackage{flushend}

\usepackage[backend=biber,style=ieee,]{biblatex}
\usepackage{geometry}
\usepackage{fancyhdr}

\fancypagestyle{firststyle}{
  \fancyhf{}
  \fancyhead[C]{2025 Asia Pacific Signal and Information Processing Association Annual Summit and Conference (APSIPA ASC)}
}

\begin{document}

\title{Spatial Audio Signal Enhancement: A Multi-output MVDR Method in The Spherical Harmonic-domain}

\author{\IEEEauthorblockN{Huawei Zhang$^1$, Jihui (Aimee) Zhang$^{2, 1, 3}$, Huiyuan (June) Sun$^{1}$, Prasanga Samarasinghe$^1$}
\IEEEauthorblockA{{$^1$Audio \& Acoustic Signal Processing Group}, {The Australian National University}, Canberra, Australia \\
{$^2$School of Electrical Engineering and Computer Science}, {The University of Queensland}, Brisbane, Australia \\
{$^3$Institute of Sound and Vibration Research}, {University of Southampton}, Southampton, U.K.\\
u7178294@anu.edu.au, jihuiaimee.zhang@uq.edu.au, u5870643@anu.edu.au, prasanga.samarasinghe@anu.edu.au}
}

\maketitle
\thispagestyle{firststyle}
\pagestyle{fancy}

\begin{abstract}
Spatial audio signal enhancement aims to reduce interfering source contributions while preserving the desired sound field with its spatial cues. 
Existing methods generally rely on impractical assumptions (e.g. accurate estimations of impractical information) or have limited applicability.   
This paper presents a spherical harmonic (SH)-domain minimum variance distortionless response (MVDR)-based spatial signal enhancer using Relative Harmonic Coefficients (ReHCs) to extract clean SH coefficients from noisy recordings in reverberant environments.
A simulation study shows the proposed method achieves lower estimation error, higher speech-distortion-ratio (SDR), and comparable noise reduction (NR) within the sweet area in a reverberant environment, compared to a beamforming-and-projection method as the baseline.
\end{abstract}

\section{Introduction}
\label{sec:introduction}
\let\thefootnote\relax\footnotetext{This work is sponsored by the Australian Research Council (ARC) Discovery Projects funding scheme with project number DE230101567. We briefly discussed the preliminary idea of this work in \cite{zhang2023directional}.}

With the rapid increase of spatial audio applications in the last decade, there has been a growing demand for spatial audio signal enhancement (also known as ambisonic-to-ambisonic separation) \cite{hu2020acoustic, lugasi2022spatial, borrelli2018denoising, herzog2020direction, hafsati2019sound, herzog2022ambisep, herzog2023ambisep, matsuda2023sound}, which separates a desired sound field from interference source contributions while preserving the desired spatial cues.
This technology acts as a fundamental step in many audio signal processing applications including sound field decomposition \cite{matsuda2023sound}, sound field recording \cite{zhang2017surround}, sound field reproduction \cite{gauthier2007adaptive, spors2013spatial, yang2022review}, and spatial active noise control \cite{zhang2018active, ma2020active, shi2023active}.

We categorize existing methods for spatial audio enhancement into three kinds: $\text{(i)}$ Beamforming-and-projection methods \cite{hu2020acoustic, lugasi2022spatial, borrelli2018denoising, herzog2020direction}, $\text{(ii)}$ Multi-channel Wiener Filter methods \cite{herzog2020direction, hafsati2019sound}, and $\text{(iii)}$ Learning-based methods \cite{herzog2022ambisep, herzog2023ambisep, matsuda2023sound}.
Beamforming-and-project methods often include at least two stages: a beamforming stage to capture the desired source signal or the desired sound pressure at a point, and a projection stage to reconstruct the desired sound field using the point-to-region acoustic transfer functions (ATFs).
While these solutions have been achieved over space via multi-point processing \cite{lugasi2022spatial} as well as spherical harmonic (SH) domain processing \cite{hu2020acoustic, borrelli2018denoising, herzog2020direction}, they make one or more impractical assumptions such as the accurate estimation of ATFs. 
Multi-channel Wiener Filter methods achieve enhancement by preserving the estimated power spectral density (PSD) matrix of the desired SH coefficients and reducing the estimated PSD matrix of interference SH coefficients.
However, the accurate estimation of the former PSD matrix is difficult.
Learning-based methods, up to today, have required a suitable dataset for pre-training.

\begin{figure}[t]
\includegraphics[width=7cm]{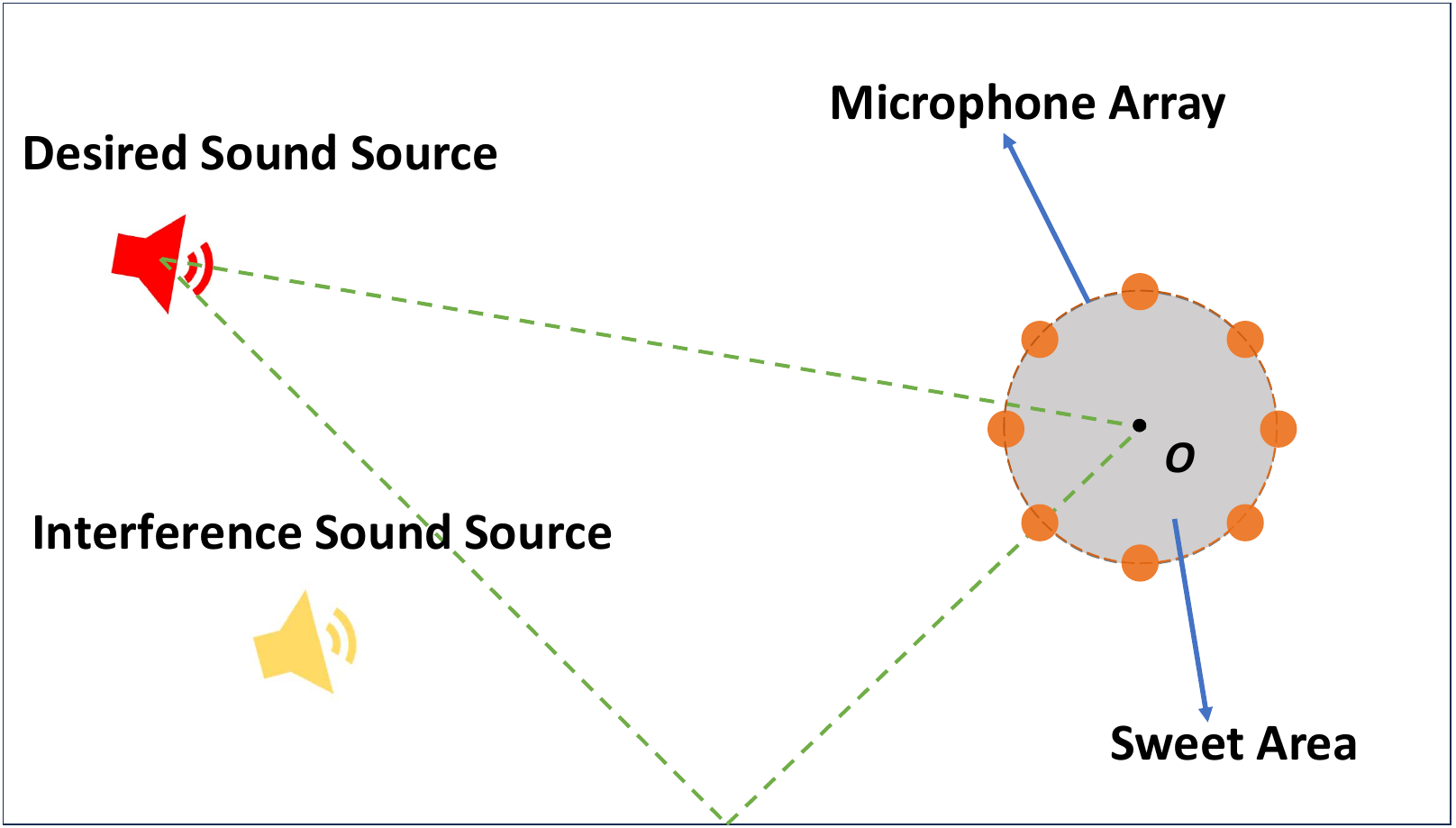}
\centering
\caption{A desired sound source (red speaker), an interference sound source (yellow speaker), and a spherical microphone array (orange solid dots) at the origin $O$ in a room environment. The grey area is the spherical sweet area. The green dash lines are examples of the acoustic paths from the desired sound source to the array. The acoustic paths from the interference sound source are omitted.}
\label{fig:layout}
\end{figure}

Relative Harmonic Coefficients (ReHCs) \cite{hu2020semi}, also named as SH-domain relative transfer functions \cite{peled2013linearly}, are denoted as the ratio between SH coefficients and the specific SH coefficient at the $0$-th order and the $0$-th mode.
These ReHCs have been used in beamformer design \cite{peled2013linearly} and spatial audio signal enhancement \cite{hu2020acoustic, herzog2020direction}.
These ReHCs can be estimated based on corresponding SH-domain PSD matrices obtained from the received microphone signals \cite{hu2020semi}.
However, the potential of ReHCs in audio enhancement has not been fully studied.

In this paper, we develop an SH-domain minimum variance distortionless response (MVDR) spatial audio signal enhancement method with multiple beamformers based on ReHCs.
Motivated by a binaural MVDR beamforming method \cite{hadad2015theoretical} and a multi-output MVDR beamforming method \cite{lugasi2022spatial}, we extend the conventional SH-domain single-output MVDR beamforming method \cite{peled2013linearly, rafaely2015fundamentals} to be a multi-output variant.
This proposed method extracts the SH coefficients due to the desired sound source from the mixed SH domain recording.
Different from the conventional single-output MVDR beamforming method (losing the spatial information) and the beamforming-and-projection methods (requiring a projection stage to reconstruct spatial information), this proposed method can directly output a set of SH coefficients representing the extracted desired sound field with its spatial cues.
Moreover, compared with relying on estimated ATFs in \cite{lugasi2022spatial}, the proposed beamforming method is developed with a more practical foreknowledge of estimated ReHCs, which is used to construct a set of spatial constraints to preserve desired SH coefficients.

\section{Problem Formulation}
\label{sec:problem}
As shown in Fig.~\ref{fig:layout}, consider a far-field scenario with a desired sound source, an interference sound source, with a spherical sweet area with the radius $r_s$ and a $Q$-microphone spherical array of the radius $r_a$ placed concentrically at the origin $O$ in a reverberated room.
The two sound sources are placed at fixed positions outside the spherical area, causing a desired sound field and an interference sound field within the sweet area, respectively.
We assume the desired sound source is uncorrelated with the interference sound source, and the interference sound source signal is stable.

Each sound field pressure within the sweet area is denoted as $x(t, k, r, \theta, \phi)$ in the time-frequency (TF) domain, where $t$ is the time frame, $k = 2 \pi f / c$ is the wavenumber, $f$ is the frequency, $c$ is the speed of sound, $r$, $\theta$, and $\phi$ are the radius, the elevation angle, and the azimuth angle, respectively. 
The sound field pressure $x(t, k, r, \theta, \phi)$ can be decomposed as
\begin{equation}
  \label{equ:pressure_divided}
  \begin{split}
    x(t, k, r, \theta, \phi) = d(t, k, r, \theta, \phi) + v(t, k, r, \theta, \phi),
  \end{split}
\end{equation} 
where $d(t, k, r, \theta, \phi)$ and $v(t, k, r, \theta, \phi)$ are the sound pressure of the desired field and the interference field, respectively.
The received microphone signal at the $q$-th microphone is denoted as
\begin{equation}
    \label{equ:pressure_sample}
    \begin{split}
        x_q(t, k) &= x(t, k, r_q, \theta_q, \phi_q) + u(t, k) \\
        &= d_q(t, k) + v_q(t, k) + u(t, k),
    \end{split}
\end{equation}
where $d_q(t, k)$, $v_q(t, k)$, $u(t, k)$ are the desired microphone signal, the interference microphone signal, and the random sensor noise, respectively, $r_q$, $\theta_q$, and $\phi_q$ are the radius, the elevation angle, and the azimuth angle, respectively, at the $q$-th microphone. 
The microphone signal $x_q(t, k)$ can be decomposed into SH coefficients to represent the mixed field as \cite{williams1999fourier, rafaely2015fundamentals}
\begin{equation}
  \label{equ:SH_dec}
  x_q (t, k) = \sum_{n=0}^{N_k} \sum_{m=-n}^n \tilde{x}_{nm}(t, k) j_n(kr_q) Y_{n m}(\theta_q, \phi_q),
\end{equation} 
where $\tilde{x}_{nm}(t, k)$ is the mixed SH coefficient representing the mixed field at the $n$-th order and the $m$-th mode, $j_n(\cdot)$ is the Spherical Bessel function of the first kind, $Y_{nm}(\cdot)$ is the SH function, and $N_k$ is the corresponding maximum order for the wavenumber $k$ defined as \cite{rafaely2015fundamentals, williams1999fourier} 
\begin{equation}
  \label{equ:SHT order wavenumber}
  N_k = \text{ceil}(kr_a) = \text{ceil}(\frac{2 \pi f}{c}r_a),
\end{equation} 
where $\mathrm{ceil}(\quad)$ is the ceiling function.
For brevity, the time frame $t$ and the wave number $k$ will be omitted in the rest of this paper.
Obtained $L = (N+1)^2$ mixed SH coefficients $\tilde{\mathbf{x}} = [\tilde{x}_{00}, \tilde{x}_{1-1}, \cdots, \tilde{x}_{nm}, \cdots, \tilde{x}_{N N}]^T$ can be divided as
\begin{equation}
  \label{equ:SH_divided}
  \begin{split}
    \tilde{\mathbf{x}} &= \text{SHT}({\mathbf{x}}) = \text{SHT}({\mathbf{d}} + {\mathbf{v}} + {\mathbf{u}}) =  \tilde{\mathbf{d}} + \tilde{\mathbf{v}} + \tilde{\mathbf{u}},
  \end{split}
\end{equation} 
where the SHT is the Spherical Harmonic Transform based on Eq.~(\ref{equ:SH_dec}) \cite{rafaely2015fundamentals, williams1999fourier}, ${\mathbf{x}} = [x_1, \cdots, x_q, \cdots, x_Q]^T$ are received microphone signals at the array, ${\mathbf{d}}$, ${\mathbf{v}}$ and ${\mathbf{u}}$ are received signals from the desired field, the interference field, and sensor noises, respectively, $\tilde{\mathbf{d}}$, $\tilde{\mathbf{v}}$ and $\tilde{\mathbf{u}}$ are SH coefficients of the desired field, the interference field, and sensor noises, respectively.

The task of this work is to extract desired SH coefficients $\tilde{\mathbf{d}}$ from mixed SH coefficients $\tilde{\mathbf{x}}$ obtained from the mixture recording $\mathbf{x}$.

\section{Proposed Multi-output MVDR method}
\label{sec:method}
In this section, we use ReHCs to develop an SH-domain multi-output MVDR beamforming method to estimate desired SH coefficients.

The proposed method is a multi-output extension of a SH-domain MVDR beamforming method, consisting of $L$ beamformers $\tilde{\mathbf{w}} = [\tilde{\mathbf{w}}_{00}^T, \tilde{\mathbf{w}}_{1-1}^T, \cdots, \tilde{\mathbf{w}}_{nm}^T, \cdots, \tilde{\mathbf{w}}_{NN}^T]^T$ as a $L^2 \times 1$ vector \cite{lugasi2022spatial, hadad2015theoretical, rafaely2015fundamentals}.
The estimated desired SH coefficient $\hat{\tilde{{d}}}_{nm}$ at the $n$-th order and the $m$-th mode is obtained with the corresponding beamformer $\tilde{\mathbf{w}}_{nm}$ as
\begin{equation}
  \label{equ:bf_sh}
  \hat{\tilde{{d}}}_{nm} = \tilde{\mathbf{w}}_{nm}^H \tilde{\mathbf{x}}.
\end{equation} 
The beamformer $\tilde{\mathbf{w}}_{nm}$ is required to preserve the desired SH coefficient and reduce SH coefficients of interference and sensor noises.
As it is difficult to obtain accurate ATFs, similar to \cite{peled2013linearly, hadad2016binaural}, we use ReHCs to develop a cost function with a spatial constraint as
\begin{equation}
  \label{equ:cost_singl_sh}
  \begin{split}
    \mathop{\operatorname{minimize}}\limits_{\tilde{\mathbf{w}}_{nm}} \quad &\tilde{\mathbf{w}}_{nm}^H \tilde{\mathbf{R}}_{\mathbf{v+u}} \tilde{\mathbf{w}}_{nm}\\
    \operatorname{subject \ to} \quad &\tilde{\mathbf{w}}_{nm}^H \tilde{\mathbf{h}} = \tilde{h}_{nm},
  \end{split}
\end{equation}
where $\tilde{\mathbf{R}}_{\mathbf{v+u}} = \tilde{\mathbf{R}}_{\mathbf{v}} + \tilde{\mathbf{R}}_{\mathbf{u}}$ is the summation of the SH-domain PSD matrices of interference $\tilde{\mathbf{R}}_{\mathbf{v}} = \mathbb{E} [\tilde{\mathbf{v}} \tilde{\mathbf{v}}^H]$ and that of sensor noises $\tilde{\mathbf{R}}_{\mathbf{u}} = \mathbb{E} [\tilde{\mathbf{u}} \tilde{\mathbf{u}}^H]$, $\mathbb{E}[\quad]$ is math expectation, $\tilde{\mathbf{h}} = [\tilde{{h}}_{00}, \tilde{{h}}_{1-1}, \cdots, \tilde{{h}}_{nm}, \cdots, \tilde{{h}}_{NN}]^T$ are ReHCs for the desired field, and $\tilde{{h}}_{nm}$ is the corresponding ReHC at the $n$-th order and the $m$-th mode.
These ReHCs $\tilde{\mathbf{h}}$ can be obtained as \cite{hu2020semi, markovich2015performance}
\begin{equation}
    \label{equ:rtf}
    \tilde{\mathbf{h}}
    = \frac{\tilde{\mathbf{d}}}{\tilde{d}_{00}} 
    =\frac{\tilde{\mathbf{R}}_{\mathbf{d}} \tilde{\mathbf{e}}_1}{\mathbf{e}_1^H \tilde{\mathbf{R}}_{\mathbf{d}} \tilde{\mathbf{e}}_1}
    \approx \frac{\hat{\tilde{\mathbf{R}}}_{\mathbf{d}+\mathbf{u}} \tilde{\mathbf{e}}_1}{\tilde{\mathbf{e}}_1^H \hat{\tilde{\mathbf{R}}}_{\mathbf{d}+\mathbf{u}} \tilde{\mathbf{e}}_1},
\end{equation}
where $\tilde{d}_{00}$ is the desired SH coefficient at the $0$-th order and the $0$-th mode, $\tilde{\mathbf{R}}_{\mathbf{d}} = \mathbb{E} [\tilde{\mathbf{d}} \tilde{\mathbf{d}}^H]$ is the SH-domain PSD matrix of desired SH coefficients, $\tilde{\mathbf{e}}_1 = [1 \quad \mathbf{0}_{1 \times (L-1)}]^T$, $\hat{\tilde{\mathbf{R}}}_{\mathbf{d}+\mathbf{u}}$ is the estimated SH-domain PSD matrix obtained from received microphone signals when only the desired source is active.

Combining cost functions as Eq.~(\ref{equ:cost_singl_sh}) in all orders and all modes, a new cost function is derived \cite{hadad2015theoretical} to design $\tilde{\mathbf{w}}$ as 
\begin{equation}
  \label{equ:cost_multi_sh}
  \begin{split}
    \mathop{\operatorname{minimize}}\limits_{\tilde{\mathbf{w}}} \quad &\tilde{\mathbf{w}}^H \tilde{\bm{\mathcal{R}}}_{\mathbf{v+u}} \tilde{\mathbf{w}}\\
    \operatorname{subject \ to} \quad & \tilde{\mathbf{C}}^H \tilde{\mathbf{w}} = \tilde{\mathbf{b}},
  \end{split}
\end{equation}
where $\tilde{\mathbf{b}} = \mathrm{conj}(\tilde{\mathbf{h}})$ is a $L \times 1$ column vector, $\mathrm{conj}(\quad)$ is complex conjugate operation, $\tilde{\bm{{\mathcal{R}}}}_{\mathbf{v+u}}$ and $\tilde{\mathbf{C}}$ are two large matrices constructed by $\tilde{\mathbf{R}}_{\mathbf{v+u}}$ and $\tilde{\mathbf{h}}$, respectively, as
\begin{equation}
  \tilde{\bm{{\mathcal{R}}}}_{\mathbf{v+u}} = 
  \begin{bmatrix}
    \tilde{\mathbf{R}}_{\mathbf{v+u}} & & & \\
     & \tilde{\mathbf{R}}_{\mathbf{v+u}} & & \\
     & & \cdots & \\
     & & & \tilde{\mathbf{R}}_{\mathbf{v+u}}
  \end{bmatrix}_{L^2 \times L^2},
\end{equation}
\begin{equation}
  \tilde{\mathbf{C}} = 
  \begin{bmatrix}
    \tilde{\mathbf{h}} & & & \\
     & \tilde{\mathbf{h}} & & \\
     & & \cdots & \\
     & & & \tilde{\mathbf{h}}
  \end{bmatrix}_{L^2 \times L}.
\end{equation}
The optimal solution of Eq.~(\ref{equ:cost_multi_sh}) can be obtained as \cite{van1988beamforming}
\begin{equation}
  \label{equ:solution_sh}
  \tilde{\mathbf{w}} = \tilde{\bm{{\mathcal{R}}}}_{\mathbf{v+u}}^{-1} \tilde{\mathbf{C}} [\tilde{\mathbf{C}}^H \tilde{\bm{{\mathcal{R}}}}_{\mathbf{v+u}}^{-1} \tilde{\mathbf{C}}]^{-1} \tilde{\mathbf{b}}.
\end{equation}
Each beamformer $\tilde{\mathbf{w}}_{nm}$ can be obtained from $\tilde{\mathbf{w}}$ directly.
After that, we can obtain an estimation $\hat{\tilde{\mathbf{d}}}$ of desired SH coefficients for all orders and all modes with Eq.~(\ref{equ:bf_sh}).

Estimated sound field pressures with spatial cues over the sweet area can be further obtained based on the Inverse Spherical Harmonic Transform (ISHT) \cite{rafaely2015fundamentals, williams1999fourier} from the estimated desired SH coefficients $\hat{\tilde{\mathbf{d}}}$.
However, spatial cues of the residual interference field may not be preserved, similar to \cite{lugasi2022spatial}.

Note the proposed method requires to obtain (1) a time interval of microphone recordings when only the desired source is active, and (2) the summation of the SH-domain PSD matrix of interference and that of sensor noises in advance.
The requirement of such a time interval can be waived by estimating the relative transfer matrix \cite{abhayapala2023generalizing} instead of ReHCs, but it is out of our scope in this paper.

\section{Simulation and Results}
\label{sec:simulation}
In this section, we evaluate the performance of the proposed method compared with a leading-edge beamforming-and-projection method \cite{lugasi2022spatial} as the baseline in a reverberant room.

The baseline is implemented by three stages:
First, we use TF-domain single-output MVDR beamforming \cite{hadad2015theoretical} to estimate a single-channel desired signal.
The beamforming requires prior knowledge of the accurate direction of arrival (DoA) and the summation of the TF-domain PSD matrix of interference and that of sensor noises.
Second, we estimate a set of ATFs based on microphone signals and our estimated desired signal, where frequency smooth operation \cite{lugasi2022spatial} has been performed with $9$ frequency bins around each target frequency bin.
Third, we implement the projection to obtain the estimated desired microphone signals.

As the output of the baseline is the estimated desired microphone signals, for a fair comparison with the proposed method, the SHT is applied to the output of the baseline to estimate desired SH coefficients.
We can then reconstruct the corresponding estimated desired field by applying the ISHT from estimated desired SH coefficients of both the proposed method and the baseline method.

\subsection{Setup}
\label{ssec:setup}

A speech signal \cite{panayotov2015librispeech} is used as the desired source at $(4.60, 4.05, 1.70)$~m.
A recorded washer-dryer noise signal \cite{reddy2019scalable} is used as the interference source at $(1.60, 1.05, 1.20)$~m. 
We set a 32-mic open spherical array at $(1.60, 4.05, 1.70)$~m, with the radius $r_a = 0.042$~m.
The array has the same size and microphone positions as the em32 Eigenmike spherical microphone array \cite{acoustics2013em32}. 
White Gaussian noises are added to the microphone array as sensor noises in terms of $35$~dB signal-sensor noise-ratio (SSNR), where the 'signal' in SSNR refers to the desired source signal.
The room size is $5\times6\times4$~m and the reverberation time is $T_{60} = 0.2$~s.
The sampling frequency is $16000$~Hz.
The radius of the sweet area is $r_s = r_a = 0.042$~m.
The speed of sound is $343$~m/s.
All room impulse responses (RIRs) are simulated with a toolbox using the image source method \cite{allen1979image, jarrett2012rigid}.

We first apply the Short-Time Fourier Transform with the frame size as $16384$ and $75\%$ overlap to the time-domain received microphone signals. 
Then we apply the SHT with the corresponding maximal order $N_k$ to each frequency bin of obtained TF-domain microphone signals.

For a fair compassion, we simulate accurate PSD matrices of interference signals and sensor noises in the TF domain and the SH domain, facilitating the baseline method and the proposed method, respectively.
The PSD matrix of interference signals is calculated based on the simulated interference source signal power and corresponding ATFs from simulated RIRs.
The PSD matrix of sensor noises is calculated by the averaged estimated PSD matrix of simulated sensor noises.

We evaluate the performance of both the proposed method and the baseline method among the common telephone bandwidth (from $300$~Hz to $3400$~Hz) \cite{hioki1998telecommunications, pulakka2013development} due to: (1) The key information of speech signals exist within the bandwidth; (2) The Spherical Bessel function of the first kind $j_n(kr)$ will achieve values close to $0$ among some higher frequency bins, resulting in the Bessel Zero problem \cite{rafaely2015fundamentals}.
Therefore, at most $3$-order SH coefficients are required based on Eq.~(\ref{equ:SHT order wavenumber}) in the following simulations.
The Bessel Zero problem can be relieved by using the spherical-shell microphone array \cite{rafaely2008spherical}.

\subsection{Performance Analysis Based on Sound Field Estimation Error}
\begin{figure}[tb]
\includegraphics[width=8cm]{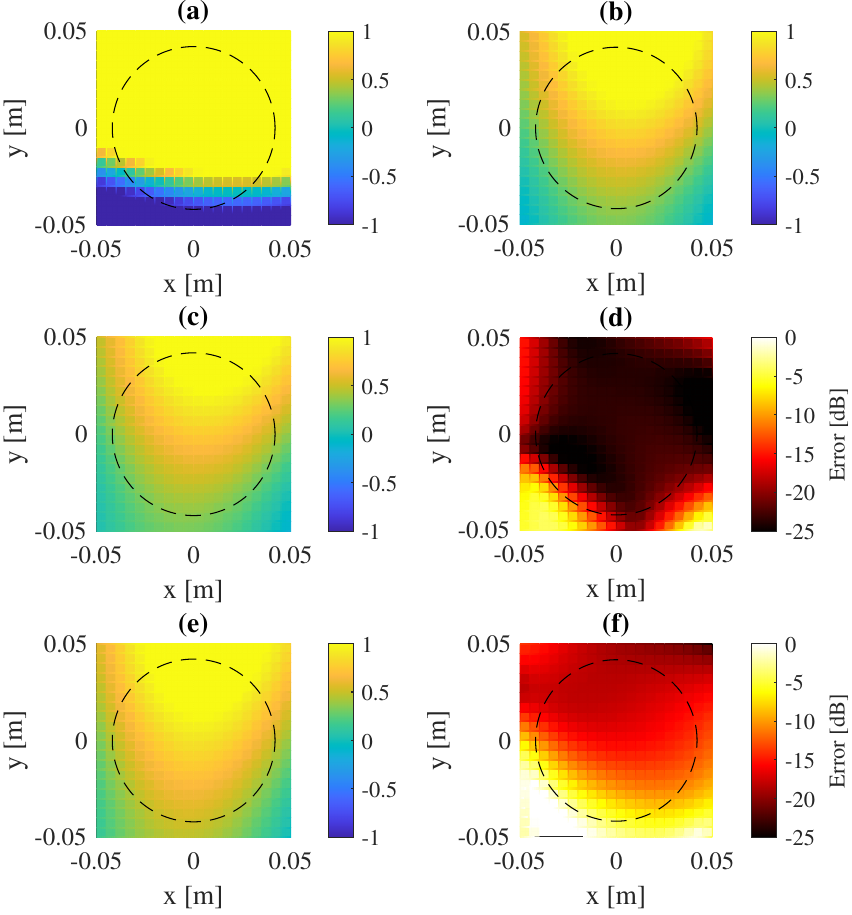}
\centering
\caption{(a) True mixed field (desired field + interference field); (b) True desired field; (c) Estimated desired field by the proposed method; (d) Estimation error for the proposed method; (e) Estimated desired field by the baseline method; (f) Estimation error for the baseline method at $1500$~Hz on the x-y plane. The black dashed circle is the border of the sweet area and the array.}
\label{fig:1500hz}
\end{figure}

In Fig.~\ref{fig:1500hz}, we compare the estimated desired sound fields and evaluate the corresponding estimation errors within the sweet area at one time frame at $1500$~Hz on the x-y plane. 
The signal-noise-ratio (SNR) is $0$~dB, where the 'signal' in SNR refers to the desired source signal and the 'noise' in SNR refers to the interference source signal.
Here, we denote the normalized square error $\epsilon(t, k)$ as the estimation error at each point represented by
\begin{equation}
  \label{equ:error}
  \epsilon(t, k) = 10 \log_{10} \frac{| {d}_{\text{point}}(t, k) - {\hat{d}}_{\text{point}}(t, k) |^2}{| {d}_{\text{point}}(t, k) |^2},
\end{equation}
where ${d}_{\text{point}}(t, k)$ and $\hat{{d}}_{\text{point}}(t, k)$ are the true desired sound field pressure and the estimated desired sound field pressure at the point, respectively.
Each sound field with corresponding estimation errors defined in Eq.~(\ref{equ:error}) is presented for $441$ observation points evenly distributed over the x-y plane.
By comparing Fig.~\ref{fig:1500hz}(b), Fig.~\ref{fig:1500hz}(c), and Fig.~\ref{fig:1500hz}(e), we observe that both the proposed method and the baseline method reconstructed a desired field similar to the true desired field within the sweet area.
Fig.~\ref{fig:1500hz}(d) and Fig.~\ref{fig:1500hz}(f) show that the proposed method achieved a lower estimation error of less than about $-15$~dB, compared with the baseline method.

\subsection{Performance Analysis over Frequency}
\label{ssec:analysis against freq}

\begin{figure}[tb]
\includegraphics[width=8.5cm]{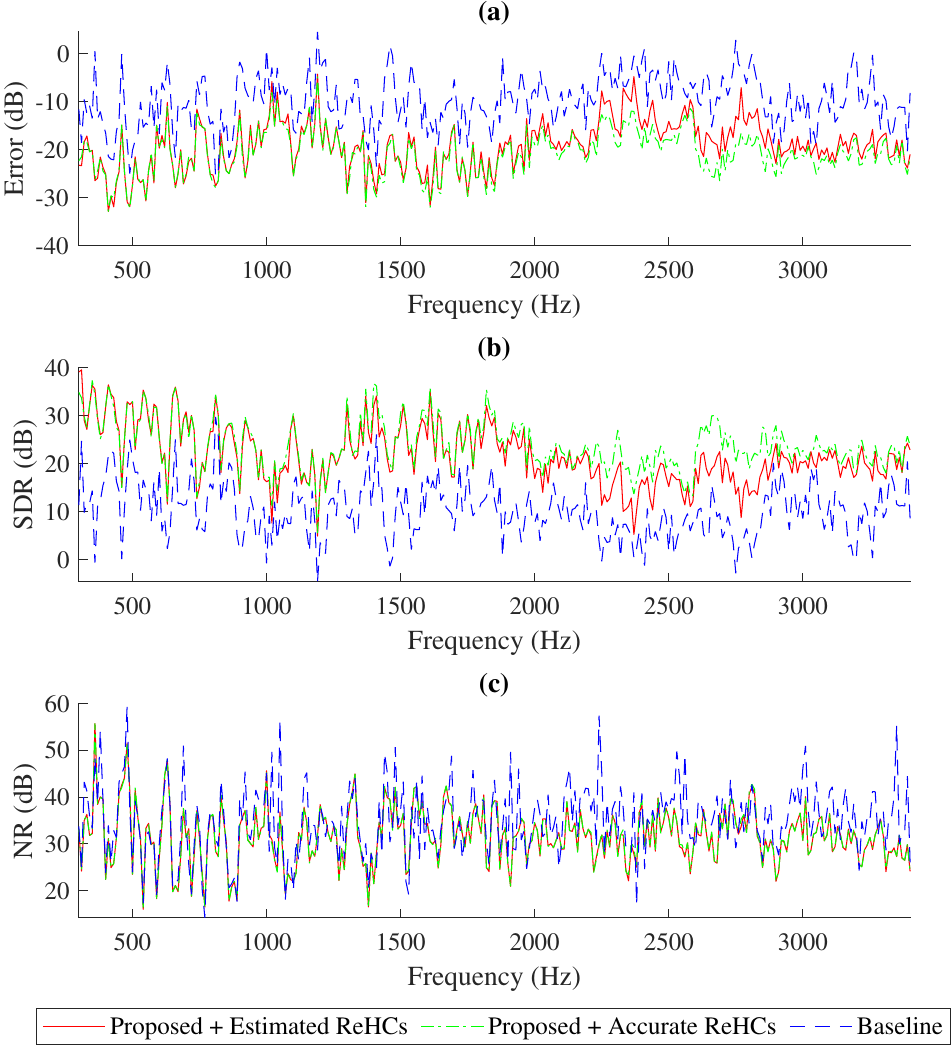}
\centering
\caption{The average of estimation error (a), SDR (b), and NR (c) of the proposed with estimated ReHCs (red), the proposed with accurate ReHCs (green), and the baseline (blue) inside the spherical sweet area among frequency bins from $300$~Hz to $3400$~Hz.}
\label{fig:error_fig}
\end{figure}

In Fig.~\ref{fig:error_fig}, we further evaluate the average of the sound field estimation error, the speech-distortion-ratio (SDR), and the noise reduction (NR) within the spherical sweet area across $15$ time frames against different signal frequencies. 
The evaluation is based on corresponding sound pressures at $107$ observation points evenly distributed over the spherical sweet area.
Here, estimation error, SDR, and NR at these $107$ observation points are defined as \cite{herzog2020direction}
\begin{equation}
  \label{equ:error_ob}
  \mathrm{Error}^\text{ob}(t, k) = 10 \log_{10} \frac{\Vert \mathbf{d}_{{s}}(t, k) - {\hat{\mathbf{d}}}_{{s}}(t, k) \Vert^2}{\Vert \mathbf{d}_{{s}}(t, k) \Vert^2},
\end{equation}
\begin{equation}
  \label{equ:sdr}
  \mathrm{SDR}^\text{ob}(t, k) = 10 \log_{10} \frac{\Vert \mathbf{d}_s(t, k) \Vert^2}{\Vert \mathbf{d}^\mathrm{res}_s(t, k) - \mathbf{d}_s(t, k) \Vert^2},
\end{equation}
\begin{equation}
  \label{equ:nr}
  \mathrm{NR}^\text{ob}(t, k) = 10 \log_{10} \frac{\Vert \mathbf{v}_s(t, k) \Vert^2}{\Vert \mathbf{v}^\mathrm{res}_s(t, k) + \mathbf{u}^\mathrm{res}_s(t, k) \Vert^2},
\end{equation}
where $\mathbf{d}_s(t, k)$, $\hat{\mathbf{d}}_{\text{s}}(t, k)$, and $\mathbf{v}_s(t, k)$ are true desired field pressures, estimated desired field pressures, and true interference field pressures, respectively, $\mathbf{d}^\mathrm{res}_s(t, k)$ and $\mathbf{v}^\mathrm{res}_s(t, k)$ are residual desired sound field pressures and residual interference sound field pressures after processing, respectively, $\mathbf{u}^\mathrm{res}_s(t, k)$ is representing the influence of residual sensor noises, at these observation points within the spherical sweet area, and $\Vert \quad \Vert$ denotes the $2$-norm operation.
As shown in Fig.~\ref{fig:error_fig}, the proposed method with estimated ReHCs achieved lower estimation error, higher SDR, and comparable NR (more than $25$~dB NR is effective enough) than the baseline method ranging from $300$~Hz to $3400$~Hz.
In detail, the proposed with estimated ReHCs achieved a lower than $-15$~dB estimation error and a higher than $15$~dB SDR at the majority of chosen frequency bins, which outperforms the baseline.

Fig.~\ref{fig:error_fig} also shows the influence of the accuracy of ReHCs. 
As a comparison, we play a segment of $0$~dB white Gaussian noise instead of the original desired source signal at the same source position, and then use the same method in Sec.~\ref{ssec:setup} to obtain another set of ReHCs as accurate ReHCs.
These ReHCs are more accurate as white Gaussian noise is more stable than speech signals at different frequency bins.
In Fig.~\ref{fig:error_fig}, we can find the proposed method with accurate ReHCs outperformed that with estimated ReHCs in terms of estimation error and SDR, and achieved comparable NR performance, which implies the performance of the proposed method is influenced by the accuracy of ReHCs estimation.

\begin{table*}[tb]
	\centering
	\caption{The averaged estimation error, SDR, and NR inside the spherical sweet area among frequency bins from $300$~Hz to $3400$~Hz across $15$ time frames against different reverberation times $T_{60}$:  (a) The proposed method; (b) The baseline method.}
	\label{tab:vary_t60}  
	\begin{tabular}{|c|c|c|c|c|c|c|c|c|c|} 
	    \hline  
		\multirow{2}{*}{} &  \multicolumn{3}{c|}{$T_{60} = 0$~s}  & \multicolumn{3}{c|}{$T_{60} = 0.2$~s} &  \multicolumn{3}{c|}{$T_{60} = 0.4$~s}  \\  
        \cline{2-10}
         & Error (dB) & SDR (dB) & NR (dB) & Error (dB) & SDR (dB) & NR (dB) & Error (dB) & SDR (dB) & NR (dB) \\  
		\hline  
        (A) The proposed & $-20.7$ & $24.0$ & $32.3$ & $-19.3$ & $21.8$ & $31.7$ & $-16.3$ & $19.3$ & $28.3$ \\
		\hline
        (B) The baseline & $-14.7$ & $14.7$ & $73.2$ & $-9.8$ & $10.2$ & $34.9$ & $-6.3$ & $7.0$ & $27.9$ \\
        \hline		
	\end{tabular}
\end{table*}

\begin{table*}[tb]
	\centering
	\caption{The averaged estimation error, SDR, and NR inside the spherical sweet area among frequency bins from $300$~Hz to $3400$~Hz across $15$ time frames against different SNR levels:  (a) The proposed method; (b) The baseline method.}
	\label{tab:vary_snr}  
	\begin{tabular}{|c|c|c|c|c|c|c|c|c|c|} 
	    \hline  
		\multirow{2}{*}{} &  \multicolumn{3}{c|}{SNR = $5$~{dB}}  & \multicolumn{3}{c|}{SNR = $0$~{dB}} &  \multicolumn{3}{c|}{SNR = $-5$~{dB}}  \\  
        \cline{2-10}
         & Error (dB) & SDR (dB) & NR (dB) & Error (dB) & SDR (dB) & NR (dB) & Error (dB) & SDR (dB) & NR (dB) \\  
		\hline  
		(A) The proposed & $-19.9$ & $21.8$ & $28.5$ & $-19.3$ & $21.8$ & $31.7$ & $-18.2$ & $21.8$ & $34.0$ \\
		\hline
        (B) The baseline & $-11.8$ & $12.0$ & $35.2$ & $-9.8$ & $10.2$ & $34.9$ & $-7.1$ & $7.6$ & $34.4$ \\
        \hline		
	\end{tabular}
\end{table*}

\subsection{Performance Analysis over Varying Reverberation Times and SNR Levels}
\label{ssec:further simulation}

Here, we evaluate the performance of both the proposed method with estimated ReHCs and the baseline method for varying reverberation times $T_{60}$ and varying SNRs, averaging within the spherical sweet area at the bandwidth for $15$ time frames.
As Sec.~\ref{ssec:analysis against freq}, three indexes are evaluated at these $107$ observation points.

Firstly, we evaluate the performance of these two methods over varying reverberation times $T_{60}$ as shown in TABLE~\ref{tab:vary_t60}.
Here, we vary the $T_{60}$ from $0$~s to $0.4$~s, while the SNR is set to be $0$~dB.
With the reverberation time $T_{60}$ increased, both methods achieved higher estimation error, lower SDR, and lower NR.
In addition, the proposed method always achieved lower estimation error, higher SDR, and comparable NR than the baseline method for varying $T_{60}$ (more than $25$~dB NR is effective enough). 

Secondly, we evaluate the performance of these two methods over varying SNR levels as shown in TABLE~\ref{tab:vary_snr}.
Here, we vary the SNR level from $5$~{dB} to $-5$~{dB}, and remain the $T_{60}$ as $0.2$~s.
TABLE~\ref{tab:vary_snr} shows that the proposed method maintained stable performance for varying SNR levels in terms of estimation error (about $-19$~dB) and SDR (about $22$~dB), always outperforming the baseline method.
By contrast, with the SNR level decreased, both estimation error and SDR achieved by the baseline method declined.
In addition, both the proposed method and the baseline method achieved comparable and stable NR (more than about $28$~dB) for varying SNR levels.

\section{Conclusion}
\label{sec:conclusion}
In this paper, we propose a spherical harmonic (SH)-domain minimum variance distortionless response (MVDR) method to estimate the desired sound field from the mixture recording at a spherical microphone array.
We use a cost function with a set of spatial constraints to extract desired SH coefficients and suppress SH coefficients of interference and sensor noises.
The field due to the desired sound source within the sweet area can be hence reconstructed.
Simulation results show that the proposed method can outperform the baseline methods within the sweet area in a reverberant room.
In the future, we plan to theoretically analyze the influence of ReHC accuracy on the performance of our proposed method.

\printbibliography

\end{document}